\documentclass[conference, a4paper]{IEEEtran}

\usepackage{amsmath}
\usepackage{amssymb}
\usepackage{latexsym}
\usepackage{verbatim}
\usepackage{subfigure}
\usepackage[final]{graphicx}
\usepackage{psfrag}

{\begin{list}               
    {$\bullet$ \hfill}{
        \setlength{\leftmargin}{\parindent}
        \setlength{\parsep}{0.04\baselineskip}
        \setlength{\itemsep}{0.5\parsep}
        \setlength{\labelwidth}{\leftmargin}
        \setlength{\labelsep}{0em}}
    }
{\end{list}}

\providecommand{\eref}[1]{\eqref{eq:#1}}  
\providecommand{\cref}[1]{Chapter~\ref{chap:#1}}
\providecommand{\sref}[1]{Section~\ref{sec:#1}}
\providecommand{\fref}[1]{Figure~\ref{fig:#1}}

\providecommand{\R}{\ensuremath{\mathbb{R}}}

\providecommand{\abs}[1]{\lvert#1\rvert}
\providecommand{\norm}[1]{\lVert#1\rVert}

\providecommand{\set}[1]{\left\{#1\right\}}

\providecommand{\bydef}{\overset{\text{def}}{=}}

\renewcommand{\vec}[1]{\ensuremath{\boldsymbol{#1}}}

\providecommand{\va}{\vec{a}}

\providecommand{\vx}{\vec{x}}

\usepackage{cite}

\usepackage{color}

\usepackage{booktabs,multirow}

\usepackage{enumitem}
\usepackage{commath}

\usepackage{algorithm, algorithmic}

\usepackage{esdiff}

\usepackage{pgf,tikz,psfrag}

\usepackage{dsfont}

\providecommand{\vxi}{\vec{\xi}}

\begin{document}
%
\title{Fundamental Limits of PhaseMax for Phase Retrieval: A Replica Analysis}

\author{\IEEEauthorblockN{Oussama Dhifallah and Yue M. Lu}
\IEEEauthorblockA{John A. Paulson School of Engineering and Applied Sciences\\
Harvard University, Cambridge, MA 02138, USA\\
{Email: oussama\_dhifallah@g.harvard.edu, yuelu@seas.harvard.edu}
}}

\maketitle

\begin{abstract}
We consider a recently proposed convex formulation, known as the PhaseMax method, for solving the phase retrieval problem. Using the replica method from statistical mechanics, we analyze the performance of PhaseMax in the high-dimensional limit. Our analysis predicts the \emph{exact} asymptotic performance of PhaseMax. In particular, we show that a sharp phase transition phenomenon takes place, with a simple analytical formula characterizing the phase transition boundary. This result shows that the oversampling ratio required by existing performance bounds in the literature can be significantly reduced. Numerical results confirm the validity of our replica analysis, showing that the theoretical predictions are in excellent agreement with the actual performance of the algorithm, even for moderate signal dimensions. 
\end{abstract}

\section{Introduction}
\label{sec:intro}

Let $\vxi \in \R^n$ be an unknown signal, and $\set{\va_i}_{1 \le i \le m}$ be a collection of sensing vectors. Given the measurements
\begin{equation}\label{eq:abs}
y_i = \abs{\va_i^T \vxi},
\end{equation}
we are interested in reconstructing $\vxi$ up to a global sign change. This is the real-valued version of the classical phase retrieval problem \cite{Gerchberg:1972jk, Fienup:82}, which has attracted much renewed interests in the signal processing community in recent years (see, \emph{e.g.}, \cite{Candes:2013xy, Jaganathan:2013zl, Waldspurger:2015rz, Netrapalli:2013qv, Candes:2015fv, WangGY:2016}). The main challenge of the phase retrieval problem comes from the nonconvex nature of the constraints \eref{abs}. Recently, a simple yet very effective convex relaxation was independently proposed by two groups of authors \cite{phmax2, phmax}. Following \cite{phmax}, we shall refer to it as the PhaseMax method, which seeks to estimate $\vxi$ via a linear programming problem:
\begin{equation}\label{eq:lp_form}
\begin{aligned}
\widehat{\vx}&=\underset{{\vx}}{\arg\,\max}~~~ {\vx}_\text{init}^{T}\,{\vx}\\	
&~~~~~~~~\text{s.t.}~~~~  \abs{\va_i^T \vx} \leq y_i, \text{ for }  1 \le i \le m.
\end{aligned}
\end{equation} 
Here, the nonconvex equality constraints in \eref{abs} have been relaxed to convex inequality constraints. The vector $\vx_\text{init}$ is an initial guess of the target vector $\vxi$. In practice, $\vx_{\text{init}}$ can be obtained if we have additional prior knowledge about $\vxi$ (\emph{e.g.}, nonnegativity) or by using a simple spectral method \cite{Netrapalli:2013qv, LuL:17}.


The performance of the PhaseMax method has been investigated in \cite{phmax,phmax2} (see also \cite{Hand:2016}), where the authors provide \emph{sufficient conditions} for PhaseMax to successfully recover the target vector $\vxi$. In this paper, we present an \emph{exact performance analysis} of the method in the high-dimensional ($n \rightarrow \infty$) limit. In particular, we show that a sharp phase transition phenomenon takes place, with a simple analytical formula characterizing the phase transition boundary.

We shall quantify the performance of PhaseMax in terms of the normalized mean squared error (NMSE), defined as $\text{NMSE}_n \bydef {\min\{\norm{\vxi - \widehat\vx}_2^2, \norm{\vxi + \widehat\vx}_2^2\}}/{\norm{\vxi}_2^2}$. The NMSE depends on two parameters: the oversampling ratio $\alpha \bydef m/n$, and the quality of the initial guess $\vx_\text{init}$, measured via the input \emph{cosine similarity}
\begin{equation}
\rho_{\text{init}} \bydef \frac{\abs{\vx_\text{init}^T \vxi}}{\norm{\vx_\text{init}}_2 \norm{\vxi}_2}.
\end{equation}
Taking values between $0$ and $1$, the parameter $\rho_\text{init}$ assesses the degree of alignment between the true signal vector $\vxi$ and the initial guess $\vx_{\text{init}}$.

\begin{figure}[t]
    \centering
    \includegraphics[width=0.28\textwidth]{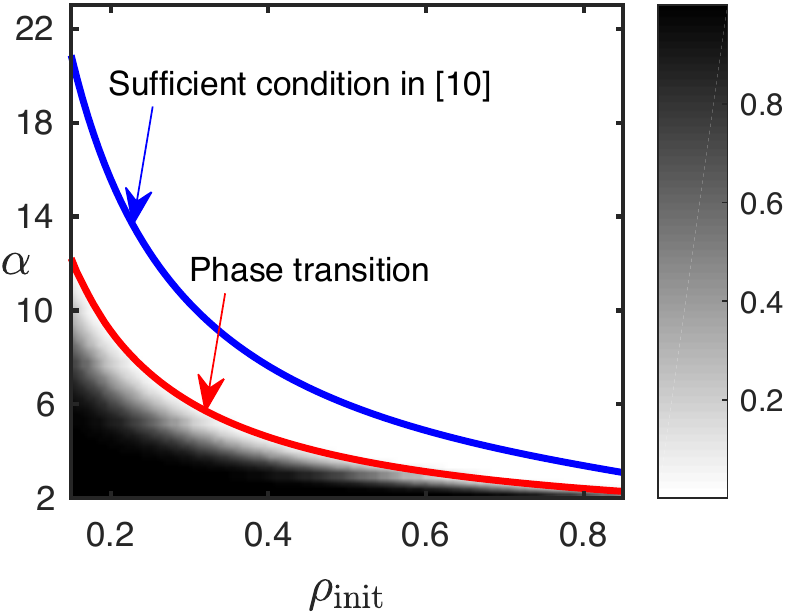}
    \caption{The normalized mean squared error (NMSE) of the PhaseMax method: theory versus simulations. The signal dimension is set to $n=1000$, and the results are averaged over $50$ independent trials. The red curve shows the predicted phase transition boundary given in \eref{pt}. Above the threshold, PhaseMax recovers the signal perfectly; below the threshold, $\text{NMSE} > 0$. The blue curve shows the sufficient condition \eref{sufficient} derived in \cite{phmax}.}
    \label{fig:phase}
\end{figure}

As the main contribution of our work, we derive the following exact asymptotic characterization of PhaseMax, under the assumption that the sensing vectors are drawn from the normal distribution:
\begin{equation}\label{eq:pt}
\mathrm{NMSE}_n \xrightarrow[]{n\to\infty} \begin{cases}
0,  &\text{if}~\frac{\pi / \alpha}{\tan(\pi / \alpha)} > 1 - \rho_\text{init}^2;\\
s(\rho_{\text{init}},\alpha), &\text{otherwise},
\end{cases} 
\end{equation} 
where $s(\rho_\text{init}, \alpha)$ is a positive function that can be computed by solving a fixed point equation (see \eref{q_fixed_point}, \eref{m_fixed_point} and \eref{s} in \sref{results_replica}.) The above expression characterizes the fundamental limits of PhaseMax: for any fixed input cosine similarity $\rho_\text{init}$, there is a critical threshold $\alpha_c(\rho_\text{init})$ such that PhaseMax perfectly recovers $\vxi$ if $\alpha > \alpha_c(\rho_{\text{init}})$, and that it fails to recover $\vxi$ if $\alpha < \alpha_c(\rho_\text{init})$. 


\fref{phase} illustrates our asymptotic characterization and compares it with results from numerical simulations. Specifically, the red curve in the figure shows the phase transition boundary $\alpha_c(\rho_{\text{init}})$ as a function of the input cosine similarity $\rho_{\text{init}}$, which can be seen to have excellent agreement with the actual performance of the algorithm. In \cite{phmax}, the authors show that PhaseMax is successful with high probability if 
\begin{equation}\label{eq:sufficient}
\alpha > \frac{2\pi}{\pi - 2 \arccos(\rho_\text{init})}.
\end{equation}
This sufficient condition is plotted as the blue curve in \fref{phase}. We note that our theoretical prediction significantly reduces the required oversampling ratio as given in \eref{sufficient} for any considered quality of the initial guess vector.

Our analysis is based on the powerful replica method \cite{Mezard:1986} from statistical mechanics. Although certain key steps of the replica method have not yet been mathematically proven, the method has been successful in the analysis of a wide-range of high-dimensional inference problems in signal and information processing (see, \emph{e.g.}, \cite{cdma_rep, cs_rep,Rangan:2012}). Some of its sharp predictions have later been proven through alternative mathematical approaches (\emph{e.g.}, \cite{Talagrand:10}). In this work, we use the replica method to derive our asymptotic predictions, and corroborate these analytical results---rigorously speaking, conjectures---via  numerical simulations.

The rest of this paper is organized as follows. After precisely laying out the various technical assumptions, we present the main results of this work in \sref{main_results}. Additional numerical results are provided in \sref{numerical} to validate our theoretical predictions. \sref{conclusion} concludes the paper. For readers interested in our replica calculations, we present some of our key derivations in the appendix, and leave the full technical details to a follow-up paper.

\section{Main Results}
\label{sec:main_results}

\subsection{Technical Assumptions}

In what follows, we first state the assumptions under which we derive our analytical predictions.

\begin{enumerate}[label={(A.\arabic*)}]
\item The sensing vectors $\set{\va_i}_{1 \le i \le m}$ are independent random vectors whose entries are i.i.d. standard normal random variables.

\item The number of measurements $m = m(n)$ with $\alpha_n = m(n) / n \rightarrow \alpha > 0$ as $n \rightarrow \infty$.

\item Both the target vector $\vxi$ and the initial guess $\vx_\text{init}$ are independent from the sensing vectors.

\item The target vector $\vxi$ has a positive cosine with the initial guess $\vx_\text{init}$.

\item $\norm{\vxi}_2 = \norm{\vx_\text{init}}_2 = \sqrt{n}$.
\end{enumerate}

Note that the last two assumptions can be made without loss of generality, since $\vxi$ and $-\vxi$ are both valid targets and thanks to the scale invariant nature of the convex optimization problem in \eref{lp_form}, respectively.

\subsection{The Boltzmann Distribution}

The first step of our replica analysis is to ``soften'' the optimization problem \eref{lp_form} via a probability distribution. To that end, we introduce the following function
\begin{equation}
\mathcal{H}(\vx) \bydef - \vx_{\text{init}}^T\,\vx - \sum\limits_{i=1}^{m} \log\big[ U ( |{\va_i^T \vxi}| - |{\va_i^T \vx}| ) \big],
\end{equation}
where $U(x)$ represents the unit-step function, \emph{i.e.}, $U(x)=1$ if $x \geq 0$ and $U(x) = 0$ otherwise. Clearly, the convex optimization problem \eref{lp_form} is equivalent to minimizing the function $\mathcal{H}$ over the variable $\vx$. Now consider the following probability distribution
\begin{align}\label{g_dist}
\mathbb{P}_{\beta}(\vx)&=\frac{1}{Z_n(\beta)} \exp(-\beta \mathcal{H}(\vx)) \nonumber\\
&=\frac{1}{Z_n(\beta)} \exp( \beta \vx_{\text{init}}^T\vx ) \prod\limits_{i=1}^{m} U ( |\va_i^T \vxi| - |\va_i^T \vx| ), 
\end{align}
where $\beta > 0$ is a fixed parameter and $Z_n(\beta)$ denotes a normalizing constant. We note that the probability distribution $\mathbb{P}_\beta(\vx)$ (often referred to as the \emph{Boltzmann distribution} in the literature) can be interpreted as a softened version of \eref{lp_form}. In particular, when $\beta = 0$, the distribution is uniform over all $\vx$ satisfying the constraints $\abs{\va_i^T \vx} \le y_i$ for $1 \le i \le m$. As we increase $\beta$, however, the distribution will become more concentrated around the optimal solution $\widehat{x}$ of \eref{lp_form}. In fact, if $\widehat{x}$ is the unique solution of \eref{lp_form}, then the distribution $\mathbb{P}_\beta(x)$ will converge to a singular distribution $\delta(x - \widehat{x})$ as $\beta \rightarrow \infty$. Therefore, to characterize the performance of the PhaseMax method, one only needs to examine the behavior of the Boltzmann distribution $\mathbb{P}_\beta(\vx)$ for sufficiently large values of $\beta$.

Of particular interest to us are the following two asymptotic moments of $\mathbb{P}_\beta(\vx)$:
\begin{align}
\vartheta_{\beta}&=\lim\limits_{n\to \infty}\frac{1}{n} \mathbb{E}\lbrace \vx^T \vxi \rbrace \label{eq:m_b}\\
q_{\beta}&=\lim\limits_{n\to \infty}\frac{1}{n} \mathbb{E}\lbrace \norm{\vx}_{2}^2 \rbrace \label{eq:q_b}
\end{align} 
where $\mathbb{E}\set{\cdot}$ denotes the expectation over the Boltzmann distribution and over the sensing vectors $\set{\va_i}_i$. Moreover, define 
\begin{equation}\label{eq:mq}
\vartheta^\ast = \lim_{\beta\to\infty} \vartheta_{\beta}\ \text{ and }\ q^\ast =\lim_{\beta\to\infty} q_{\beta}.
\end{equation}
Since the Boltzmann distribution $\mathbb{P}_\beta(\vx)$ will be concentrated on the global optimal solution $\widehat{\vx}$ of \eref{lp_form} as $\beta \to \infty$, the value of $\vartheta^\ast$ reveals the normalized inner product between the optimal solution $\widehat\vx$ and the true signal vector $\vxi$, \emph{i.e.}, $\vartheta^\ast=\lim_{n\to\infty}\frac{1}{n} \mathbb{E}\{\widehat\vx^T \vxi\}$. Similarly, the value of $q^\ast$ is equal to the normalized squared norm of $\widehat\vx$, \emph{i.e.}, $q^\ast=\lim_{n\to\infty}\frac{1}{n} \mathbb{E} \lbrace \norm{\widehat\vx}_{2}^2 \rbrace$. It follows that the asymptotic NMSE can be computed as
\begin{align}
 \mathrm{NMSE}&=\lim\limits_{n\to+\infty}\frac{\mathbb{E} \min\{\norm{\vxi - \widehat\vx}_2^2, \norm{\vxi + \widehat\vx}_2^2\}}{\norm{\vxi}_2^2}\nonumber\\
 &=q^\ast-2\abs{\vartheta^\ast}+1,\label{eq:nmsehd}
\end{align}
where in reaching \eref{nmsehd} we have used the assumption that $\norm{\vxi}_2 = \sqrt{n}$. Thus, the task of analyzing the asymptotic performance of PhaseMax boils down to calculating the values of $q^\ast$ and $\vartheta^\ast$, which we do next by using the replica method.

\subsection{Asymptotic Predictions via the Replica Method}
\label{sec:results_replica}

Much of the information about the Boltzmann distribution $\mathbb{P}_\beta(\vx)$, in particular, the moments as in \eref{m_b} and \eref{q_b}, can be obtained from its log-partition function, defined as
\[
f(\beta) \bydef \lim_{n\to\infty} \frac{-1}{\beta n} \mathbb{E}\big\lbrace \log Z_n(\beta) \big\rbrace.
\]
The challenge here is to compute the partition function $Z_n(\beta)$, which involves a high-dimensional integration. And this is where the replica method \cite{Mezard:1986} comes in. Using this method, we can calculate $f(\beta)$ for all $\beta > 0$. In particular, its limit as $\beta \to \infty$ can be derived as
\begin{align}\label{eq:fed_4}
f({\infty}) &=\underset{\substack{\vartheta,q,\chi\\
   \widehat{\vartheta},\widehat{q},\widehat{\chi}}}{\operatorname{extr}}\Bigg[ \frac{\alpha}{2\pi\chi} \Bigg\lbrace -2 \sqrt{q-\vartheta^2}+ (1+q+2 \vartheta) A_1 \nonumber\\ &+ (1+q-2 \vartheta)A_2 
-4 \vartheta A_3 \Bigg\rbrace -\frac{1}{2}\widehat{Q}q+\frac{1}{2}\widehat{\chi}\chi \nonumber\\
&+\widehat{\vartheta} \vartheta-\frac{\widehat{\vartheta}^2+1+2\widehat{\vartheta} \rho_{\text{init}}}{2\widehat{Q}} -\frac{\widehat{\chi}}{2\widehat{Q}} \Bigg],
\end{align}
where $A_1=\mathrm{arctan}\big( \frac{q+\vartheta}{\sqrt{q-\vartheta^2}} \big)$, $A_2=\mathrm{arctan}\big( \frac{q-\vartheta}{\sqrt{q-\vartheta^2}} \big)$ and $A_3=\mathrm{arctan}\big( \frac{\vartheta}{\sqrt{q-\vartheta^2}} \big)$ are three functions, and $\mathrm{extr}_\mathcal{X} \lbrace g(X) \rbrace$ denotes the extremization of a function $g$ over a set of variables $\mathcal{X}$. Here, $\mathcal{X}$ includes six scalar variables $\vartheta$, $q$, $\chi$, $\widehat{\vartheta}$, $\widehat{q}$ and $\widehat{\chi}$. To streamline our presentation, we postpone the details of our replica calculations leading to \eref{fed_4} to the appendix.

To solve the extremization problem in \eref{fed_4}, we set the gradient with respect to the variables to zero, which leads to a set of nonlinear saddle point equations. After some further simplifications, we can eliminate the variables $\vartheta$, $\chi$, $\widehat{\vartheta}$, $\widehat{q}$ and $\widehat{\chi}$ and just need to study a simple fixed-point equation:
\begin{equation}\label{eq:q_fixed_point}
\begin{aligned}
q &= h(q)\\
 &\bydef 1 + \frac{\pi}{2\alpha c}\big[w^2(q) - {\rho^{-2}_\text{init}}{(\vartheta - w(q))^2} - 1\big],
\end{aligned}
\end{equation}
where $c = \tan(\pi/\alpha)/2$ is a constant,
\begin{equation}\label{eq:m_fixed_point}
\vartheta = \sqrt{q - c^2(1-q)^2}
\end{equation}
and
\begin{equation}
w(q) = 1 - \tfrac{2\alpha}{\pi} \arctan\Big( \frac{c \abs{1-q}}{1 + \sqrt{q - c^2(1-q)^2}}\Big).
\end{equation}
The solution to the above equations then gives us the key parameters of interest $\vartheta^\ast$ and $q^\ast$ as defined in \eref{mq}, from which we can compute the asymptotic NMSE by using \eref{nmsehd}.

We observe that $q = 1$ is always a fixed point of \eref{q_fixed_point}. However, for any fixed $\rho_\text{init}$ and when we reduce the oversampling ratio $\alpha$ to below a threshold, a second fixed point emerges and the original solution $q = 1$ becomes unstable. This is indeed the origin of the phase transition. To locate the phase transition boundary, we study the stability of the solution $q = 1$. Specifically, by the definition of $h(q)$, we can verify that $\od{h}{q}\sVert[1]_{q =1} \equiv 1$ and
\[
\dod[2]{h}{q}\sVert[3]_{q = 1} = \frac{\alpha c}{2\pi} - \frac{ (\pi/2- \alpha c)^2}{2\pi \alpha c \rho^2_\text{init}} - 1/4.
\]
Thus, the solution $q = 1$ becomes unstable (\emph{i.e.} a phase transition happens) when
\[
\rho^2_\text{init} < 1 - \frac{\pi/\alpha}{\tan(\pi/\alpha)},
\]
which is exactly when $\od[2]{h}{q}\sVert[1]_{q = 1}$ changes its sign. Finally, the function $s(\rho_\text{init}, \alpha)$ in \eref{pt} can be obtained as
\begin{equation}\label{eq:s}
s(\rho_\text{init}, \alpha) = q^\ast - 2\abs{\vartheta^\ast} + 1,
\end{equation}
where $q^\ast$ is the stable solution of $\eref{q_fixed_point}$ and $\vartheta^\ast$ is given by \eref{m_fixed_point}.

\section{Numerical Results}
\label{sec:numerical}

In this section, we present additional numerical results to verify our analytical predictions found through the replica method. In all of our experiments, we solve the convex optimization problem \eref{lp_form} using the approach presented in \cite{fasta} where the signal dimension is set to $n=1000$. The results are also averaged over $50$ independent Monte Carlo trials. 

\begin{figure}[t]
    \centering
    \subfigure[]{\label{fig:nmse_1}
    \includegraphics[width=0.47\linewidth]{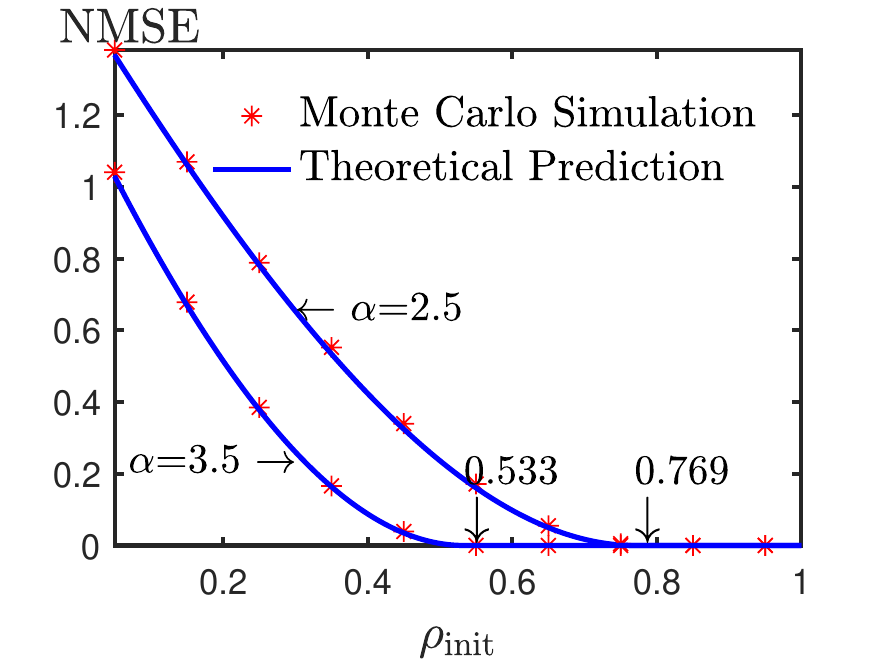}
    }
    \subfigure[]{\label{fig:nmse_2}
        \includegraphics[width=0.47\linewidth]{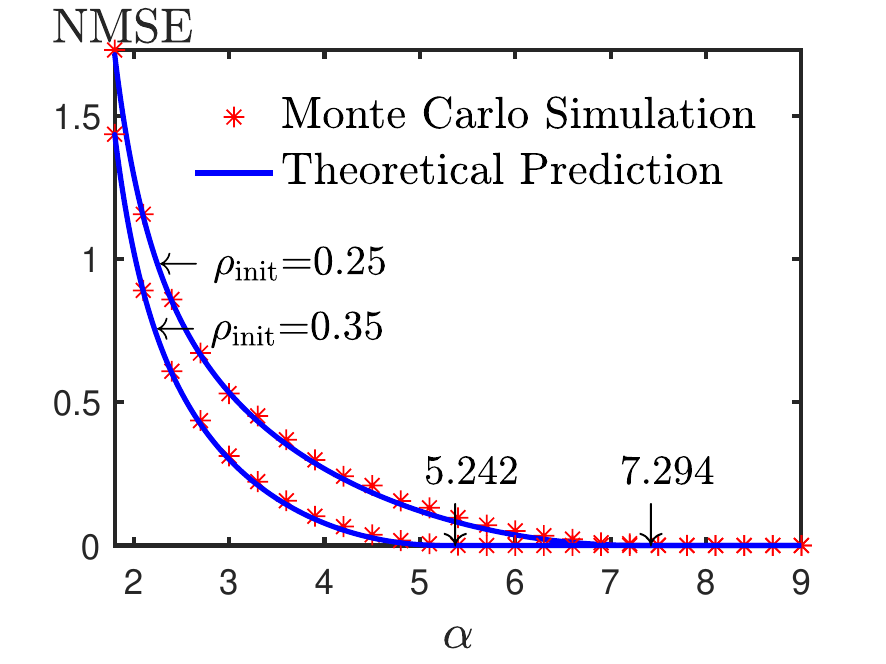}}

    \caption{Analytical predictions v.s. numerical simulations. (a) The NMSE as a function of the input cosine similarity, for two different values of the oversampling ratio; (b) The NMSE as a function of the oversampling ratio, for two different values of $\rho_\text{init}$. In both cases, the theoretical results obtained by the replica analysis can accurately predict the actual performance of the PhaseMax method.}
    \label{fig1}
\end{figure}

Our first simulation example, shown in \fref{nmse_1}, studies the performance of our analytical prediction of the NMSE [see \eref{s}] as a function of the input cosine similarity $\rho_{\text{init}}$ for two different values of the oversampling ratio: $\alpha = 2.5$ and $\alpha = 3.5$, respectively. As seen from the figure, the theoretical prediction obtained by the replica method is in excellent agreement with the experimental results obtained by numerically solving the convex optimization problem \eref{lp_form}. The results also validate our theoretical prediction of the phase transition points: the critical input cosine similarity corresponding to each value of $\alpha$ is $\rho_{\text{init}}(\alpha=2.5)=0.769$ and $\rho_{\text{init}}(\alpha=3.5)=0.533$.

A different example is shown in \fref{nmse_2}, where we examine the performance of our analytical prediction of the NMSE as a function of the oversampling ratio $\alpha$ for two different values of the input cosine similarity: $\rho_{\text{init}} = 0.25$ and $\rho_\text{init} = 0.35$, respectively. Again, as seen from the figure, our theoretical results can accurately predict the actual performance of the algorithm.

\section{Conclusion}
\label{sec:conclusion}

We presented in this paper an exact characterization of the performance of the PhaseMax method for phase retrieval. Our replica analysis leads to an analytical formula for the asymptotic normalized MSE of the estimate given by PhaseMax in the high-dimensional limit. It also reveals a sharp phase transition phenomenon: for PhaseMax to succeed, the oversampling ratio must be above a critical threshold, given as a function of the input cosine similarity. Simulation results confirm the validity of our theoretical predictions. They also show that our theoretical results significantly reduce the required oversampling ratio given by an existing sufficient condition in the literature.

\appendix
\section{Technical Details}
\label{app_a}
This appendix provides a sketch of our derivations leading to \eref{fed_4}. To start, we write the partition function $Z_n(\beta)$ as 
\begin{align}
\hspace{-2.5mm} Z_n(\beta)=\int \exp\left( \beta \vx_{\text{init}}^T\vx \right) \prod\limits_{i=1}^{m} U \Big\lbrace |\va_i^T \vxi| - |\va_i^T \vx| \Big\rbrace ~d\vx.\nonumber
\end{align}
Using the replica trick \cite{Mezard:1986}, the \emph{free energy density} for any given parameter $\beta$ can be expressed as follows 
\begin{align}
f(\beta)&=\lim_{n\to\infty} \frac{-1}{\beta n} \mathbb{E}\Big\lbrace \log(Z_n(\beta)) \Big\rbrace\nonumber\\
&=\lim_{n\to\infty} \lim_{\ell\to 0} \frac{-1}{\beta \ell n} \log \Big\lbrace \mathbb{E}(Z^\ell_n(\beta)) \Big\rbrace,
\end{align}
where the expectation is over the random measurement vectors $\lbrace \va_i,~1\leq i \leq m \rbrace$. To determine the expression of the free energy density, we first need to compute an analytical expression of $\mathbb{E}(Z^\ell_n(\beta))$. To this end, we write the expression of $\mathbb{E}(Z^\ell_n(\beta))$ as follows
\begin{align}\label{znbeta2}
&\mathbb{E}(Z^\ell_n(\beta))=\int \exp\left(\sum\limits_{a=1}^{\ell}\beta \vx_{\text{init}}^T\vx_a \right) \times \nonumber\\
 &\qquad\left[ \mathbb{E}\left\lbrace  \prod\limits_{a=1}^{\ell} ~U \Big\lbrace \frac{|\va^T \vxi| }{\sqrt{n}} - \frac{|\va^T \vx_a| }{\sqrt{n}} \Big\rbrace \right\rbrace \right]^m  \prod\limits_{a=1}^{\ell} d\vx_a,
\end{align}
where $\vx_a$ denotes the $a${th} replica signal and where the expectation is over the random vector $\va$ which is normally distributed with zero mean and covariance matrix ${\bf I}_n$.

Define the following variables $\vartheta_a=n^{-1}\vxi^T \vx_a$, $r_a=n^{-1}\vx_a^T \vx_a$ and $q_{ab}=n^{-1}\vx_a^T \vx_b$, for all $a,b \in \lbrace 1,..,\ell \rbrace$. The expression of $\mathbb{E}(Z^\ell_n(\beta))$ given in (\ref{znbeta2}) can be rewritten as 
\begin{align}\label{znbeta4}
&\mathbb{E}(Z^\ell_n(\beta)) =\int \exp \Bigg( n \Big\lbrace \alpha \log \left[  \mathcal{G}(\lbrace \vartheta_a, r_a, q_{ab} \rbrace) \right] \nonumber\\
 &\quad+ \mathcal{I}(\lbrace \vartheta_a, r_a, q_{ab} \rbrace) \Big\rbrace  \Bigg)  \prod\limits_{a=1}^{\ell} d\lbrace r_a \rbrace d\lbrace \vartheta_a \rbrace \prod\limits_{a<b}^{} d\lbrace q_{ab} \rbrace,
\end{align}
where $\alpha=\frac{m}{n}$ denotes the oversampling ratio and where the function $\mathcal{G}$ can be expressed as follows
\begin{align}
\mathcal{G}(\left\lbrace \vartheta_a,r_a,q_{ab} \right\rbrace) = \mathbb{E}\left\lbrace  \prod\limits_{a=1}^{\ell} ~U \Big\lbrace |u_0| - |u_a| \Big\rbrace \right\rbrace,
\end{align} 
with $u_0=\va^T \vxi/\sqrt{n}$ and $u_a=\va^T\vx_a/\sqrt{n}$, for all $a$. Furthermore, using a result in large deviation theory known as the Gartner-Ellis theorem \cite{ldt}, the rate function $\mathcal{I}$ can be expressed as the Fenchel--Legendre transform of a cumulant generating function. Specifically, the rate function $\mathcal{I}$ can be expressed as follows
\begin{align}
&\mathcal{I}(\lbrace \vartheta_a, r_a, q_{ab} \rbrace)=\underset{\tilde{\vartheta}_a,\tilde{r}_a,\tilde{q}_{ab}}{\operatorname{extr}}\Bigg( -\sum\limits_{a=1}^{\ell} \tilde{\vartheta}_a \vartheta_a-\sum\limits_{a=1}^{\ell} \tilde{r}_a r_a\nonumber\\
&\qquad-\sum\limits_{1\leq a < b \leq \ell}^{} \tilde{q}_{ab} q_{ab}+ \lambda(  \lbrace \tilde{\vartheta}_a,\tilde{r}_a,\tilde{q}_{ab} \rbrace ) \Bigg),
\end{align}
where the function $\mathcal{\lambda}$ represents the cumulant generating function and is given by
\begin{align}
&\lambda(\lbrace \tilde{\vartheta}_a,\tilde{r}_a,\tilde{q}_{ab} \rbrace) = \lim_{n\to \infty}\frac{1}{n} \log\Bigg[ \int \exp \Bigg( \sum\limits_{a=1}^{\ell} \tilde{\vartheta}_a \vxi^T \vx_a \nonumber\\
&+ \sum\limits_{a=1}^{\ell} \tilde{r}_a \vx^T_a \vx_a + \sum\limits_{a < b}^{} \tilde{q}_{ab} \vx^T_a \vx_b +\sum\limits_{a=1}^{\ell}\beta \vx_{\text{init}}^T\vx_a \Bigg) \prod\limits_{a=1}^{\ell} d{\vx_a} \Bigg].\nonumber
\end{align}
We use the replica symmetry (RS) ansatz where it is assumed that, when the dimension $n$ is sufficiently large, the integration in (\ref{znbeta4}) is dominated by the configurations satisfying the following particular property: $\vartheta_a=\vartheta$, $\tilde{\vartheta}_a=\tilde{\vartheta}$, $r_a=r$, $\tilde{r}_a=\tilde{r}$, $q_{ab}=q$ and $\tilde{q}_{ab}=\tilde{q}$, for all $a,b \in \lbrace 1,2,..,\ell \rbrace$. This leads to the following representation of the random variables $u_0$ and $\lbrace u_a, 1 \le a \le \ell \rbrace$:
\begin{equation}
\begin{aligned}
u_0&=\sqrt{1-\vartheta^2/q}s_0+\sqrt{\vartheta^2/q} t\\
u_a&=\sqrt{r-q}s_a+\sqrt{q} t
\end{aligned}
\end{equation}
where $s_0$, $t$ and $\lbrace s_a, 1\leq a \leq n \rbrace$ are i.i.d. Gaussian random variables with zero mean and unit variance. Using the introduced representation of the random variables $u_0$ and $u_a$, the free energy density can be rewritten as follows
\begin{align}\label{fed_3}
 &f(\beta)=\underset{\substack{\vartheta,q,r\\
   \tilde{\vartheta},\tilde{q},\tilde{r}}}{\operatorname{extr}}\Bigg\lbrace  -\frac{1}{2\beta}\tilde{q}q+\frac{1}{\beta}\tilde{r}r+\frac{1}{\beta}\tilde{\vartheta}\vartheta+\frac{1}{2\beta}\log(\tilde{q}-2\tilde{r}) \nonumber\\
&-\frac{\alpha}{\beta} \mathbb{E}_{s_0,t} \left\lbrace \log \left[ \Phi\left( \frac{\left| u_0 \right|-\sqrt{q}t}{\sqrt{r-q}} \right)-\Phi\left( \frac{-\left| u_0 \right|-\sqrt{q}t}{\sqrt{r-q}} \right) \right] \right\rbrace \nonumber\\
&-\frac{\tilde{\vartheta}^2+\beta^2+2\tilde{\vartheta}\beta \rho_\text{init}}{2\beta(\tilde{q}-2\tilde{r})} -\frac{\tilde{q}}{2\beta(\tilde{q}-2\tilde{r})} \Bigg\rbrace,
\end{align}
where the expectation is over the random variables $s_0$ and $t$ and where $\Phi$ denotes the cumulative distribution function of the standard normal distribution. Recall that our objective is to study the behavior of the free energy density as the parameter $\beta$ tends to infinity. When $\beta$ goes to $+\infty$, the only non-trivial solution of the extremization problem given in (\ref{fed_3}) occurs when $\chi=\beta(r-q)$ is finite. Assume that $\widehat{\chi}=\beta^{-2} \tilde{q}$, $\widehat{\vartheta}=\beta^{-1} \tilde{\vartheta}$ and $\widehat{Q}=\beta^{-1} (\tilde{q}-2\tilde{r})$. Under the RS assumption and using the Laplace method, the free energy density, as $\beta$ tends to infinity, can then be expressed as in \eref{fed_4}.

\IEEEtriggeratref{11}

\bibliographystyle{IEEEtran}
\bibliography{reference,refs}

\begin{thebibliography}{10}
\providecommand{\url}[1]{#1}
\csname url@samestyle\endcsname
\providecommand{\newblock}{\relax}
\providecommand{\bibinfo}[2]{#2}
\providecommand{\BIBentrySTDinterwordspacing}{\spaceskip=0pt\relax}
\providecommand{\BIBentryALTinterwordstretchfactor}{4}
\providecommand{\BIBentryALTinterwordspacing}{\spaceskip=\fontdimen2\font plus
\BIBentryALTinterwordstretchfactor\fontdimen3\font minus
  \fontdimen4\font\relax}
\providecommand{\BIBforeignlanguage}[2]{{%
\expandafter\ifx\csname l@#1\endcsname\relax
\typeout{** WARNING: IEEEtran.bst: No hyphenation pattern has been}%
\typeout{** loaded for the language `#1'. Using the pattern for}%
\typeout{** the default language instead.}%
\else
\language=\csname l@#1\endcsname
\fi
#2}}
\providecommand{\BIBdecl}{\relax}
\BIBdecl

\bibitem{Gerchberg:1972jk}
R.~W. Gerchberg, ``A practical algorithm for the determination of phase from
  image and diffraction plane pictures,'' \emph{Optik}, vol.~35, p. 237, 1972.

\bibitem{Fienup:82}
J.~R. Fienup, ``Phase retrieval algorithms: a comparison,'' \emph{Applied
  Optics}, vol.~21, no.~15, pp. 2758--2769, 1982.

\bibitem{Candes:2013xy}
E.~J. Candes, T.~Strohmer, and V.~Voroninski, ``Phaselift: {Exact} and stable
  signal recovery from magnitude measurements via convex programming,''
  \emph{Communications on Pure and Applied Mathematics}, vol.~66, no.~8, pp.
  1241--1274, 2013.

\bibitem{Jaganathan:2013zl}
K.~Jaganathan, S.~Oymak, and B.~Hassibi, ``Sparse phase retrieval: {Convex}
  algorithms and limitations,'' in \emph{Information {Theory} {Proceedings}
  ({ISIT}), 2013 {IEEE} {International} {Symposium} on}.\hskip 1em plus 0.5em
  minus 0.4em\relax IEEE, 2013, pp. 1022--1026.

\bibitem{Waldspurger:2015rz}
I.~Waldspurger, A.~d'Aspremont, and S.~Mallat, ``Phase recovery, maxcut and
  complex semidefinite programming,'' \emph{Mathematical Programming}, vol.
  149, no. 1-2, pp. 47--81, 2015.

\bibitem{Netrapalli:2013qv}
P.~Netrapalli, P.~Jain, and S.~Sanghavi, ``Phase retrieval using alternating
  minimization,'' in \emph{Advances in {Neural} {Information} {Processing}
  {Systems}}, 2013, pp. 2796--2804.

\bibitem{Candes:2015fv}
E.~J. Candes, X.~Li, and M.~Soltanolkotabi, ``Phase retrieval via {Wirtinger}
  flow: {Theory} and algorithms,'' \emph{Information Theory, IEEE Transactions
  on}, vol.~61, no.~4, pp. 1985--2007, 2015.

\bibitem{WangGY:2016}
G.~Wang, G.~B. Giannakis, and Y.~C. Eldar, ``Solving {Systems} of {Random}
  {Quadratic} {Equations} via {Truncated} {Amplitude} {Flow},''
  \emph{arXiv:1605.08285}, May 2016.

\bibitem{phmax2}
\BIBentryALTinterwordspacing
S.~Bahmani and J.~Romberg, ``{Phase Retrieval Meets Statistical Learning
  Theory: {A} Flexible Convex Relaxation},'' \emph{CoRR}, vol. abs/1610.04210,
  2016. [Online]. Available: \url{http://arxiv.org/abs/1610.04210}
\BIBentrySTDinterwordspacing

\bibitem{phmax}
\BIBentryALTinterwordspacing
T.~Goldstein and C.~Studer, ``{{PhaseMax}: Convex Phase Retrieval via Basis
  Pursuit},'' \emph{CoRR}, vol. abs/1610.07531, 2016. [Online]. Available:
  \url{http://arxiv.org/abs/1610.07531}
\BIBentrySTDinterwordspacing

\bibitem{LuL:17}
\BIBentryALTinterwordspacing
Y.~M. Lu and G.~Li, ``Phase transitions of spectral initialization for
  high-dimensional nonconvex estimation,'' \emph{arXiv:1702.06435 [cs.IT]},
  2017. [Online]. Available: \url{https://arxiv.org/abs/1702.06435}
\BIBentrySTDinterwordspacing

\bibitem{Hand:2016}
P.~Hand and V.~Voroninski, ``Corruption {Robust} {Phase} {Retrieval} via
  {Linear} {Programming},'' \emph{arXiv:1612.03547 [cs, math]}, Dec. 2016.

\bibitem{Mezard:1986}
M.~Mezard, G.~Parisi, and M.~Virasoro, \emph{Spin {Glass} {Theory} and
  {Beyond}: {An} {Introduction} to the {Replica} {Method} and {Its}
  {Applications}}, ser. World {Scientific} {Lecture} {Notes} in
  {Physics}.\hskip 1em plus 0.5em minus 0.4em\relax World Scientific, Nov.
  1986, vol.~9.

\bibitem{cdma_rep}
T.~Tanaka, ``{A statistical-mechanics approach to large-system analysis of CDMA
  multiuser detectors},'' \emph{IEEE Transactions on Information Theory},
  vol.~48, no.~11, pp. 2888--2910, Nov 2002.

\bibitem{cs_rep}
Y.~Kabashima, T.~Wadayama, and T.~Tanaka, ``{A typical reconstruction limit for
  compressed sensing based on $\ell_p$ -norm minimization},'' \emph{Journal of
  Statistical Mechanics: Theory and Experiment}, vol. 2009, no.~09, p. L09003,
  2009.

\bibitem{Rangan:2012}
S.~Rangan, A.~K. Fletcher, and V.~K. Goyal, ``Asymptotic analysis of {MAP}
  estimation via the replica method and applications to compressed sensing,''
  \emph{Information Theory, {IEEE} Transactions on}, vol.~58, no.~3, pp.
  1902--1923, 2012.

\bibitem{Talagrand:10}
M.~Talagrand, \emph{Mean Field Models for Spin Glasses}.\hskip 1em plus 0.5em
  minus 0.4em\relax Springer, 2010, vol. 1 and 2.

\bibitem{fasta}
\BIBentryALTinterwordspacing
T.~Goldstein, C.~Studer, and R.~Baraniuk, ``A field guide to forward-backward
  splitting with a {FASTA} implementation,'' \emph{arXiv eprint}, vol.
  abs/1411.3406, 2014. [Online]. Available:
  \url{http://arxiv.org/abs/1411.3406}
\BIBentrySTDinterwordspacing

\bibitem{ldt}
H.~Touchette, ``The large deviation approach to statistical mechanics,''
  \emph{Physics Reports}, vol. 478, no.~1, pp. 1 -- 69, 2009.

\end{thebibliography}

\end{document}